\documentclass[
reprint,
superscriptaddress,
amsmath,amssymb,
prstab,
]{revtex4-2}
\usepackage{float}
\usepackage{placeins}
\usepackage{enumerate}
\usepackage{csquotes}
\usepackage{graphicx}% Include figure files
\usepackage[dvipsnames]{xcolor}% colors for text
\usepackage{amsmath}
\usepackage{subcaption}
\usepackage{siunitx}
\DeclareSIUnit\bar{bar}
\DeclareSIUnit\var{VAR}
\usepackage{tablefootnote}
\graphicspath{ {./images/} }
\usepackage{lipsum}                     % Dummytext
\usepackage{xargs}                      % Use more than one optional parameter in a new commands
\usepackage[pdftex,dvipsnames]{xcolor}  % Coloured text etc.
\begin{document}
\preprint{AIP/123-QED}
\title{Design of a High-Average-Power Ferroelectric Phase Shifter}
\author{Ilan \surname{Ben-Zvi}}
\email{Ilan.Ben-Zvi@StonyBrook.edu}
\affiliation{Physics and Astronomy Department, Stony Brook University, NY USA}
\author{Alick \surname{Macpherson}}
\email{alick.macpherson@cern.ch}
\affiliation{CERN, CH-1211 Geneva, Switzerland}
\author{Samuel \surname{Smith}}
\email{Samuel.jack.smith@cern.ch}
\affiliation{CERN, CH-1211 Geneva, Switzerland}
\date{\today}
\begin{abstract}
This paper describes performance of a voltage controlled phase shifter designed for a high average-power and a high figure of merit. The device is a reflection-type, resonant phase shifter that utilizes ferroelectric capacitors and impedance matching to the port impedance. With state-of-the-art ferroelectric materials, the device achieves a high Figure of Merit of nearly a thousand degrees phase shift per dB insertion loss at 800 MHz,  along with sub-microsecond response. This phase shifter is capable of handling average power up to a megawatt. Its application  is expected to enhance the electrical efficiency of particle accelerators, reducing both capital and operating costs.
\end{abstract}

\maketitle

%\pagebreak

\section{Introduction}\label{section1}
Phase shifters are essential components in microwave electronics, including RF systems for particle accelerators. A voltage-controlled phase shifter is characterized by parameters such as center frequency,   phase shift range,  bandwidth, insertion loss, response time and power handling capability.  Depending on the application, certain parameters become critical.

This paper describes the performance of a voltage-controlled phase shifter designed for high average power and a high Figure of Merit ($FoM$), defined as the degrees of phase shift over its full range, normalized to the insertion loss in dB. The device is a reflection-type, resonant phase shifter that utilizes ferroelectric capacitors and impedance matching to the port impedance.  With state-of-the-art ferroelectric materials, the device achieves a high $FoM$ of nearly 1,000 degrees of phase shift per dB of insertion loss at 800 MHz, along with a sub-microsecond response time. This phase shifter is capable of handling average power levels up to a megawatt.  Its application is expected to enhance the electrical efficiency of particle accelerators, reducing both capital and operational costs.

A transmission line loaded with a ferroelectric material
serves as a transmission type phase shifter. Such
a phase shifter has the advantage of a broad frequency
response. The permittivity controls the phase advance
through this line. Losses result both from mismatch at
the input port, $S11$ losses, and from transmission through
the line, $S21$ losses. Since the characteristic impedance of
the loaded line change with permittivity, the $S11$ losses
vary with the phase shift. Reflection-type phase shifters, which terminate a transmission line with a permittivity-controlled variable reactance, eliminate $S21$ losses and minimize $S11$ losses, resulting in a high $FoM$. However, their bandwidth is narrower compared to transmission-type phase shifters. 

 Previous studies on ferroelectric phase shifters have demonstrated limited power handling capabilities. The phase shifter described here is designed to improve the  average power  capability and the $FoM$. 

 The reflected power must be separated from the forward power by additional components, however in many applications this separation is not required. For example, a frequency tuner of a cavity is a reflection-type phase shifter, acting by reflecting power flowing out of the cavity right back.

The idea of using the properties of ferroelectric materials to produce a phase shifter has been tried in various configurations. Cohn and Eikenberg \cite{Cohn2003} present the analysis, construction and performance of compact surface wave ferroelectric phase shifters for VHF and UHF frequencies. The anticipated power handling capability is approximately 40 kW peak power. The measured performance at 200 MHz was a 360 degrees phase shift, with an insertion loss of 3 to 4 dB, leading to a $FoM$ of about 90 to 100 degrees per dB,  a number that is typical for commercial phase shifters.  Kazakov et. al. \cite{kazakov2010fast} presented a design of a magic tee based 1.29 GHz  120 degrees phase shifter aimed at a pulsed power of 500 kW and average power of 4 kW. A prototype was tested at low power,  demonstrating a pulse response of 90 ns rise time, inferring the ferroelectric material temporal response of 30 ns for a 77 degrees phase change. The insertion loss measured for that device varied between about 1.9 dB at no bias to about 2.2 dB for a 6 kV bias, corresponding to approximately 120 degrees phase change. The $FoM$ is then about 60 degrees per dB. 
A reflection type and resonant phase shifter based on a ferroelectric capacitor in a thin film circuit has been described by Vendik 
\cite{vendik2007insertion}, stating an expected $FoM$ of 200, and providing experimental data at 8-9 GHz with a measured $FoM$ of 90 degrees per dB.
Another phase shifter, using a pair of reflection type phase shifters used with a hybrid coupler is described by Kim et al, \cite{kim2002wide}. 
All the work cited above is limited to a low average RF power.
The subject of this work is a reflection-type, resonant phase shifter capable of very high power at a high FoM. It is based on a novel structure of ferroelectric wafers described in \cite{ben2024conceptual} and \cite{BenZvi2025Detailed}. As an application that requires the ultimate performance of a phase shifter: fast response time, high $FoM$ and very high average power capability, consider a high-power frequency converter \cite{BenZvi2025HighpowerFFM}. This application is aimed at enabling a phase-locked high-efficiency magnetron as an RF power source for accelerators, allowing a large savings in capital investment and electrical power consumption. 
 
\section{The Ferroelectric Material}\label{section2_2}

High average-power operation requires low-loss ferroelectric material and a material configuration that readily removes dissipated heat.

Currently the ferroelectric of choice is a BST ceramic (BaTiO3/SrTiO3-Mg) \cite{kanareykin2005low}, which  has a room temperature of permittivity  of $\sim\epsilon=160$, and is a low loss ceramic, with a very low frequency dependent loss tangent. The typical measured loss tangent value is $\delta = 3.4 \times 10^{-4}$ at \SI{80}{\mega \hertz}, and a frequency scaling as $\sim \omega ^{0.783}$ between \SI{80}{\mega \hertz} and \SI{400}{\mega \hertz}, depending on material  temperature  has been reported \cite{kanareykin2005low}. 

 The geometry for high power heat removal is realized by a flat wafer format, potentially stacked as a number of layers, thus permitting a reduced bias voltage for a given polarizing electric field (small wafer thickness) and improving the thermal conductivity for cooling the wafer (larger wafer surface area). In addition, reference \cite{ ben2024conceptual} demonstrated that an annular wafer geometry reduces the parasitic inductance and resistance at the capacitor-spacer transition. 

Power dissipation in the ferroelectric material dominates the technical challenges when designing high-power devices. The performance of the bare ferroelectric material can be described by a Material Figure of Merit $F$, given by  
\begin{equation}
    F=\frac{\Delta \epsilon}{2 \delta \epsilon_c}
    \label{eq4}
\end{equation}
where $\Delta  \epsilon=|\epsilon_2 -\epsilon_1|$, and $\epsilon_c$ is a central value for the relative permittivity, approximated by $\epsilon_c \approx \sqrt{\epsilon_1 \epsilon_2 }$. 
This definition of $F$ is used in \cite{ben2024conceptual}, and is related to Vendik's Commutation Quality Factor $K$ \cite{vendik2007insertion} through
\begin{equation}
    F=\frac{1}{2}\sqrt{K}
    \label{Vendik K}    
\end{equation}

For the aforementioned BST ceramic,  this $F$ is temperature dependent with a  peak at about \SI{50}{\degreeCelsius}. At this temperature, with permittivity end-states defined by a sustainable bias electric field of up to 8 MV/m,  $\epsilon_1=96, \epsilon_2=130$, and the frequency dependent loss tangent at \SI{400}{\mega \hertz} is  $\delta =9.5 \times 10^{-4}$. This in turn implies a $F=156$.

To determine the minimal aspect ratio of an individual wafer, with a one-sided surface area of $A$, assume a power dissipation of $P_W$ in the two-wafer series-connected stack. Given a power dissipation $P_W$,  an   average temperature rise $\Delta T$ results:
\begin{equation}
    \begin{aligned}[c]
        \Delta T&= \frac{P_W g}{12K_{FE}A}
    \end{aligned}
    \label{T_Aopt}    
\end{equation}
Here, $K_{FE}$ is the thermal conductivity  of the ferroelectric material \cite{kanareykin2005low}.

\section{The Phase Shifter}\label{Methodology}

\begin{figure}[tb]
    \centering
    \includegraphics[width=0.95\columnwidth]{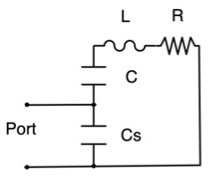}
    \caption{\label{Circuit} The equivalent circuit of a capacitive coupled reflection phase shifter. $C$ represents the ferroelectric capacitor, $C_s$  a series coupling capacitor, $L$ the inductance and $R$ the combined resistance of the inductor and capacitor.} 
\end{figure}

The tuner analysed in \cite{ BenZvi2025Detailed} employs a reflection-type phase shifter, comprising a resonant circuit with a ferroelectric capacitor and a transmission line. This demanding tuner design is designed to handle a large reactive power of the order of 400 kVAR. 
This paper explores the performance of a high-power phase shifter using such a resonant circuit in a reflection mode, as shown in Figure \ref{Circuit}. 

For  optimum tuner performance, one may adjust the coupling of the resonator to the impedance $Z_0$ of the port, given by the transformer turn ratio $N$

 \begin{equation}
    N \equiv \frac{C}{C_s + C}
    \label{Transformer}
\end{equation}

The impedance of the resonant circuit at the port is given by

\begin{equation}
   Z_r(\xi)= \frac{\frac{R}{Q}QN^2}{1+jQ\frac{C_s}{C+C_s}\frac{\Delta\epsilon}{2\epsilon}\xi}
    \label{Zr}
\end{equation}
Here $\xi$ varies from $-1$ to $+1$.

 The quality factor $Q$ of the resonator is given by:

\begin{equation}
   Q= \frac{1}{\omega \frac{C C_s}{C+C_s} (R+\frac{\delta}{\omega C})}
    \label{Q}
\end{equation}

For the sake of simplicity, assume for the time being that the losses in the resonant circuit are due just to the average ferroelectric loss tangent $\delta$. This approximation is reasonable at high microwave frequencies.

With this approximation,

\begin{equation}
   Q\frac{C_s}{C+C_s}= \frac{1}{\delta}
    \label{Qsimple}
\end{equation}

The impedance of the resonator simplifies to

\begin{equation}
   Z_r(\xi)= \frac{\frac{R}{Q}QN^2}{1+jF\xi}
    \label{Zr Simple}
\end{equation}
$F$ is the material's Figure of Merit defined in the previous section.
 Now define $D$ as
\begin{equation}
   D\equiv\frac{\frac{R}{Q}QN^2}{F Z_0}
    \label{Define D}
\end{equation}
$D$ is a parameter which represents the strength of the coupling of the resonator to the transmission line.
Then
\begin{equation}
   Z_r(\xi)= \frac{F D Z_0}{1+jF\xi}
    \label{Zr with D}=\frac{F D Z_0(1-j\xi F)}{1+(\xi F)^2}
\end{equation}

The impedance of the resonant circuit leads directly to the reflection coefficient $\Gamma$, which depends on the permeability setting, the material FoM $F$ and the coupling strength $D$. 

In the following, approximations  will be used to obtain useful analytic expressions. In these approximations it is assumed that losses in the ferroelectric alone are responsible for the power dissipation, and that the material losses are small. However, the results presented in Figures 1 through 6 were obtained with exact calculations using Maple \cite{maple2019}, where the conductor and ferroelectric losses are taken into account.

For a compact representation, define the following:

$A\equiv FD$, $B\equiv 1+(\xi F)^2$, $C\equiv \xi F^2 D$
 then
\begin{equation}
 \begin{aligned}[c]
    \Gamma=  & \frac{A^2-B^2+C^2-2jBC)}{(A+B)^2+C^2}
       \end{aligned}
    \label{Gamma}
\end{equation}

Using the reflection coefficient, the phase shift and insertion loss can be worked out.
The tangent of the phase shift $\theta$ is given by:
\begin{equation}
   tan(\theta)= -\frac{2\xi F^2 D(1+(\xi F)^2)}{(FD)^2-(1+(\xi F)^2)^2+(\xi F^2 D)^2}
    \label{tan theta}
\end{equation}

The power insertion loss is

\begin{equation}
 \begin{aligned}[c]
    \left| \Gamma \right|^2=  & \frac{(A^2-B^2+C^2)^2+(2BC)^2}{((A+B)^2+C^2)^2}
       \end{aligned}
    \label{power loss}
\end{equation}

First, evaluate the phase shift at the extremal range points ( $\xi=\pm 1$ ) using $F^2 \gg 1$. The evaluation depends if the complex number representing the reflection coefficient is in the first quadrant (if $D>1$ or in the second quadrant (for $D<1$).

\begin{equation}
    \begin{aligned}[c]
    &tan(\theta)= \pm \frac{2D}{D^2-1}    \hspace{0.3cm}\text{while $D>1$}\\
   &tan(\pi -\theta)= \pm \frac{2D}{1-D^2} \hspace{0.3cm}\text{while $D<1$}\\
   \end{aligned}
    \label{tan theta range}
\end{equation}
\begin{figure}[H]
    \centering
    \includegraphics[width=0.95\columnwidth]{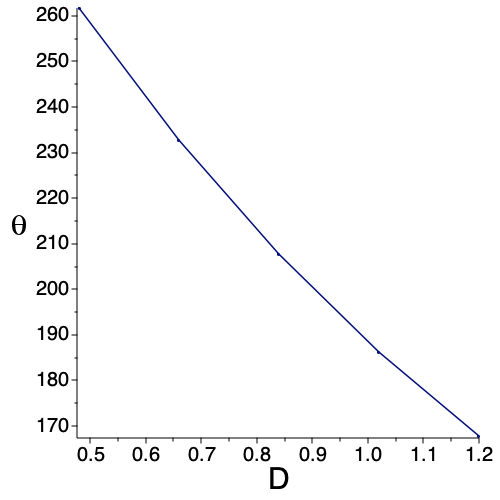}
    \caption{\label{phase bs D} The phase shift range in degrees as a function of the coupling parameter D.} 
\end{figure}

Equation \ref{tan theta range} demonstrates that the range is not a function of $F$ as long as $F$ is large. For a strong coupling, $D>1$ the phase shifting range is smaller than 180 degrees. At $D=1$ (critical coupling) the range is exactly 180 degrees. The choice of $D$ determines the phase shifting range, as can be seen in Figure \ref{phase bs D}. This figure has been produced without the above approximations.

The insertion loss of the phase shifter measured in $dB$ is important too, particularly for high-average power applications. It is possible to provide an analytical approximate value for the insertion loss at the edges of the phase shift range and in its middle (at $\xi=0$). As mentioned earlier, the power dissipation in the copper resonator are neglected. Given that the dominant problem in high average power ferroelectric devices is the power dissipation in the ferroelectric material, these expressions are useful 

At $\xi=0$ one obtains:
\begin{equation}
  \left |\Gamma \right |= \frac{FD-1}{FD+1}\approx 1-\frac{2}{FD}
    \label{abs S11 mid}
\end{equation}

The insertion loss in $dB$ is

\begin{equation}
  L_i= 20 log_{10}(\left |\Gamma \right |) \approx \frac{20}{2.303}\frac{2}{FD}\approx 17.37\frac{1}{FD}
    \label{dB mid}
\end{equation}

For the special case of $D=1$ the $FoM$ in degrees per dB is
\begin{equation}
  FoM= 10.36 \cdot F
    \label{dB FoM mid}
\end{equation}

\begin{figure}[H]
    \centering
    \includegraphics[width=0.95\columnwidth]{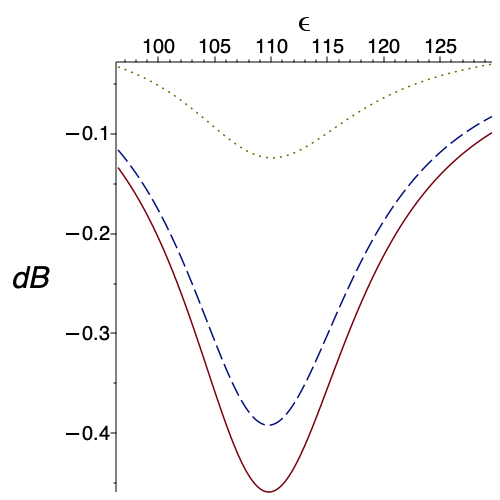}
    \caption{\label{dB vs eps} The insertion loss in dB as a function of the permittivity $\epsilon$ at 800 MHz and D=0.6. The dotted line corresponds to contribution from copper loss, the dashed line to the contribution of the ferroelectric loss tangent and the solid line to all loss mechanisms included.} 
\end{figure}

\begin{figure}[H]
    \centering
    \includegraphics[width=0.95\columnwidth]{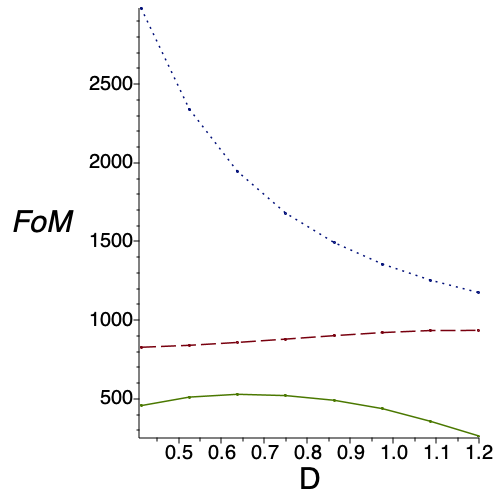}
    \caption{\label{FoM vs D} The FoM in degrees per dB at 800 MHz as a function of the coupling parameter $D$. The dotted line corresponds to insertion loss at the extreme ends of the tuning range, the dashed line to the average insertion loss and the solid line to the worst insertion loss.} 
\end{figure}

The $FoM$ can be also evaluated where the insertion loss is taken at the extreme ends of the range, useful for a phase shifter that stays most of the time at 0 or 180 degrees phase shift. The result is
\begin{equation}
  FoM= 10.36 \cdot 2 F
    \label{dB FoM edge}
\end{equation}
This result has been obtained earlier by Vendik \cite{vendik2007insertion}.

The insertion loss as a function of permittivity is shown in Figure \ref{dB vs eps} where the relative contributions of dissipation in the copper and in the ferroelectric are illustrated for a phase shifter operating at 800 MHz. Clearly the approximation made in the previous chapter where the loss in the copper was neglected is reasonable at this (and higher) frequencies. 

Depending on the application requirements, it is informative to provide the  $FoM$ at various situations: Worst insertion loss,  extreme end-points ($\xi=\pm 1$) or average insertion loss.  This is shown in Figure \ref{FoM vs D}.

\begin{figure}[H]
    \centering
    \includegraphics[width=0.95\columnwidth]{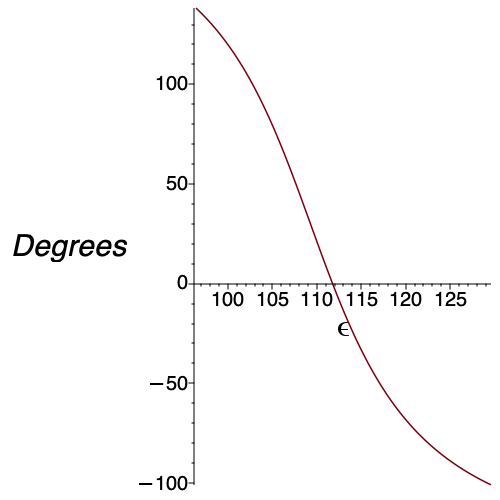}
    \caption{\label{Phase vs epsilon} The phase shift in degrees at D=0.6 as a function of permittivity. The total phase range is 239 degrees.} 
\end{figure}
The $FoM$ using the insertion loss at the edges of the tuning range (180 degrees commutation) is twice as high as for the insertion loss at mid-range (worst case insertion loss). For the application of a phase-shifter to convert RF power efficiently from one frequency to another, where the load is tolerant to a small, high frequency amplitude modulation, the average insertion loss is most appropriate.

For designing phase shifters at various frequencies, one must consider the frequency dependence of material property $F$. The phase shifting range is not a function of frequency as can be seen in equation \ref{tan theta range}, showing that  $\theta$ does not a depend on $F$. However, the insertion loss, as in equation \ref{abs S11 mid}, is inversely proportional to $F$, thus the $FoM$ is proportional to $F$.

Using the mathematics software Maple \cite{maple2019}, it is straightforward to evaluate the quantities of interest like the phase shift and insertion loss for a wide range of parameters while avoiding some of the approximations made in the previous section.

Given the impedance of the resonator of equation \ref{Zr}, the reflection coefficient and thus the phase shift and insertion loss can be derived for realistic conditions (including copper losses) as a function of frequency, material properties or coupling strength.

With the phase shift range and insertion loss given, it is now possible to observe the $FoM$ dependence on the coupling parameter. Given that the insertion loss is not constant but changes with the permittivity (or phase shift), one must consider the intended application of the phase shifter.

\begin{itemize}
    \item 180 degrees phase commutator. The FoM operates at the extreme ends of the tuning range, where the highest FoM is obtained. Potential application: Digital phase shifters.
    \item Arbitrary variable phase, small amplitude modulation tolerated. FoM uses the phase-averaged insertion loss. Potential application: Frequency conversion of a magnetron driving high Q RF cavities.
    \item Arbitrary variable phase with no amplitude modulation. FoM uses the worst insertion loss figure, amplitude modulation is used to maintain constant output. Potential application: Distribution of a high RF power source to several accelerator cavities with phase and amplitude control. 
\end{itemize}

 These three possible choices for determining the $FoM$ are presented in  Figure \ref{FoM vs D}, with the $FoM$ plotted as a function of the coupling parameter $D$. The best value of $D$ can be selected depending on the intended application of the phase shifter.

 At this point a choice can be made for the coupling parameter $D$. The choice of $D=0.6$ will be adopted in the following. Another item of interest is the phase shift in degrees as a function of permittivity. This is shown for $D=0.6$ in Figure \ref{Phase vs epsilon}

\begin{figure}[H]
    \centering
    \includegraphics[width=0.95\columnwidth]{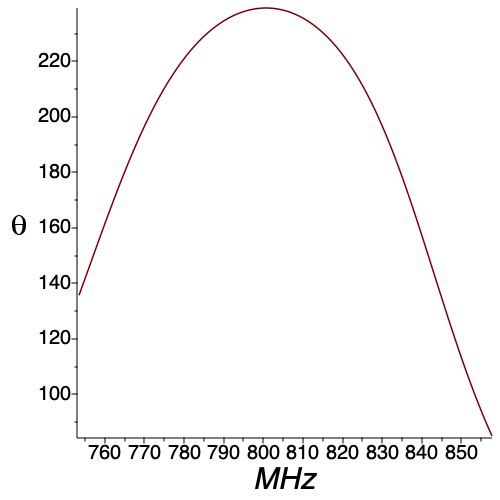}
    \caption{\label{Phase vs frequency} The phase shift $\theta$ in degrees at D=0.6 as a function of frequency. The phase range at 800 MHz is 239 degrees.} 
\end{figure}

Finally, the instantaneous bandwidth of the phase shifter may be of interest. This is shown in Figure \ref{Phase vs frequency} for a center frequency of 800 MHz.

\begin{figure}[H]
    \centering
\includegraphics[width=0.95\columnwidth]{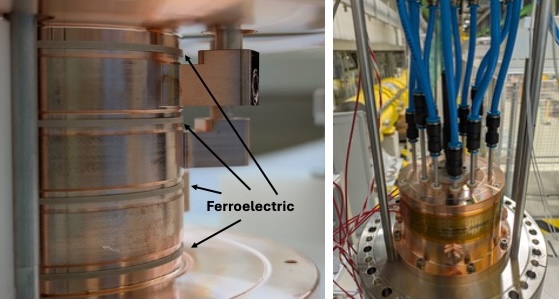}
    \caption{\label{TDD1} {High power CERN phase shifter to be tested as a tuner of a 400 MHz cavity. Left: Interior of the resonant circuit, showing the ferroelectric wafers between cooling spacers. Right: The resonant phase shifter with liquid cooling lines.} }
\end{figure}

\section{SUMMARY}\label{Summary}

This paper presents a theoretical analysis of a high-power ferroelectric fast phase shifter  with a Figure of Merit of approximately 1000 degrees per dB.  
A reflection-type resonant circuit is exemplified by the tuner  described in \cite{ben2024conceptual}, \cite{BenZvi2025Detailed}, which is designed for 400 kW reactive power.  This tuner  has been constructed at CERN (Figure \ref{TDD1}) and is undergoing experimental performance evaluation. 
The phase-shifter's high performance makes it suitable for applications such as frequency conversion \cite{BenZvi2025HighpowerFFM} in RF systems, enabling significant cost and energy savings in particle accelerators.

\begin{acknowledgments}
This work is in part supported through the  Innovate for Sustainable Accelerating Systems  (iSAS) programme funded through the European Commission’s Horizon Europe Research and Innovation programme under Grant Agreement n°101131435. 
I.B-Z. acknowledges support under the CERN Visiting Scientist program.

\end{acknowledgments}

\bibliography{bib}

\end{document}